\documentclass[manuscript]{acmart}

\PassOptionsToPackage{hyphens}{url}\usepackage{hyperref}

\definecolor{phrases}{gray}{0.5}

\usepackage{subcaption}

\AtBeginDocument{%
  \providecommand\BibTeX{{%
    \normalfont B\kern-0.5em{\scshape i\kern-0.25em b}\kern-0.8em\TeX}}}

\setcopyright{ccbysa}
\copyrightyear{2021}
\acmYear{2021}
\copyrightyear{2020}
\acmDOI{}

\acmConference[recsyslabs 2021]{recsyslabs Technical Report}{2021}{}
\acmBooktitle{recsyslabs Technical Report 2021-X001}
\acmPrice{}
\acmISBN{}

\begin{document}

\title{Personalization, Privacy, and Me}



\author{Reshma Narayanan Kutty}
\affiliation{%
  \institution{\texttt{recsyslabs}}
  \city{}
  \country{}
}
\affiliation{%
  \institution{University College Dublin}
  \city{}
  \country{Ireland}
}
\email{reshma@recsyslabs.com}


\author{Claudia Orellana-Rodriguez}
\affiliation{%
  \institution{\texttt{recsyslabs}}
  \city{}
  \country{}
}
\affiliation{%
  \institution{University College Dublin}
  \city{}
  \country{Ireland}
}
\email{claudia@recsyslabs.com}


\author{Igor Brigadir}
\affiliation{%
  \institution{\texttt{recsyslabs}}
  \city{}
  \country{}
}
\affiliation{%
  \institution{University College Dublin}
  \city{}
  \country{Ireland}
}
\email{igor@recsyslabs.com}


\author{Ernesto Diaz-Aviles}
\affiliation{%
  \institution{\texttt{recsyslabs}}
  \city{}
  \country{}
}
\affiliation{%
  \institution{University College Dublin}
  \city{}
  \country{Ireland}
}
\email{ernesto@recsyslabs.com}

\renewcommand{\shortauthors}{R. Narayanan Kutty et al.}

\begin{abstract}
  News recommendation and personalization is not a solved problem. People are growing concerned of their data being collected in excess in the name of personalization and the usage of it for purposes other than the ones they would think reasonable. Our experience in building personalization products for publishers while adhering to safeguard user privacy led us to investigate more on the user perspective of privacy and personalization. We conducted a survey to explore people's experience with personalization and privacy and the viewpoints of different age groups. In this paper, we share our major findings with publishers and the community that can inform algorithmic design and implementation of the next generation of news recommender systems, which must put the human at its core and reach a balance between personalization experiences and privacy to reap the benefits of both. 
\end{abstract}




\maketitle

\section{Introduction}
\label{sec:intro}
Personalization today comes at a huge price, including giving out unnecessary personal information, which is at risk of being exposed in case of data breaches and other unauthorized access. With businesses increasingly incorporating personalization technology, the data and usage have started getting more attention by users and regulators. With Google and other big players banning third-party cookies that target personal information from their browsers, the awareness and need to consider user privacy is more than ever for news outlets and digital publishers.

Data breaches and information leak from companies have made privacy a major concern in recent years. Laws, like General Data Protection Regulation (GDPR)~\cite{GDPR} and California Consumer Privacy Act (CCPA)~\cite{ccpa}, are in place to strictly govern companies for their usage of customer data. Even when user identities are anonymized, their activities can identify them~\cite{atfop}, revealing protected categories such as race, gender, sexual orientation, or religion.

To design better personalization products we need to understand how growing privacy concerns affect reader experiences~\cite{10.1145/3314183.3323672}. Whether, for example, different age groups take or do not take action to protect their privacy online. How enjoyable it is to receive personalized content online if privacy is not (fully) respected~\cite{10.1145/3313831.3376415}, and how aware people are of how much of their data are collected and/or necessary in exchange for a personalized experience~\cite{nyt_privacy}.

To strengthen our study, outside our quantitative research, we also conducted a survey to understand more about consumer behavior, identify user pain points, and learn about the current state of users accessing digital content in relationship with their privacy concerns. The purpose of the study is to understand user awareness about data privacy and the transparency in the usage of user data.

In the rest of the paper we report the main lessons learned with the aim to help the community, both in academia and industry, to reach a balance between personalization experiences and privacy protection for the next generation of recommender systems we envision. 

\section{Results}
\label{sec:results}
The survey was shared across social media, including LinkedIn and Twitter, and garnered 270 responses over a period of 1 month during July 2020. 84\% of the respondents were Millennials and Generation Z (Gen Z), and 16\% were over the age of 35.

Survey recruitment strategy: The survey mainly targets digital users who access content online and are used to giving out personal information in exchange for content personalization. Social media platform was used to share the survey across even though no personal information was captured through the survey.

The results of our recent survey shed light on what readers today really care about when they access media online. The main insights are as follows.

\subsection*{Personalization remains a favorite among younger age groups}
\label{sub:personalization-remains-a-favorite-among-younger-age-groups}
\begin{figure}[t!]
  \centering
  \includegraphics[width=0.618\linewidth]{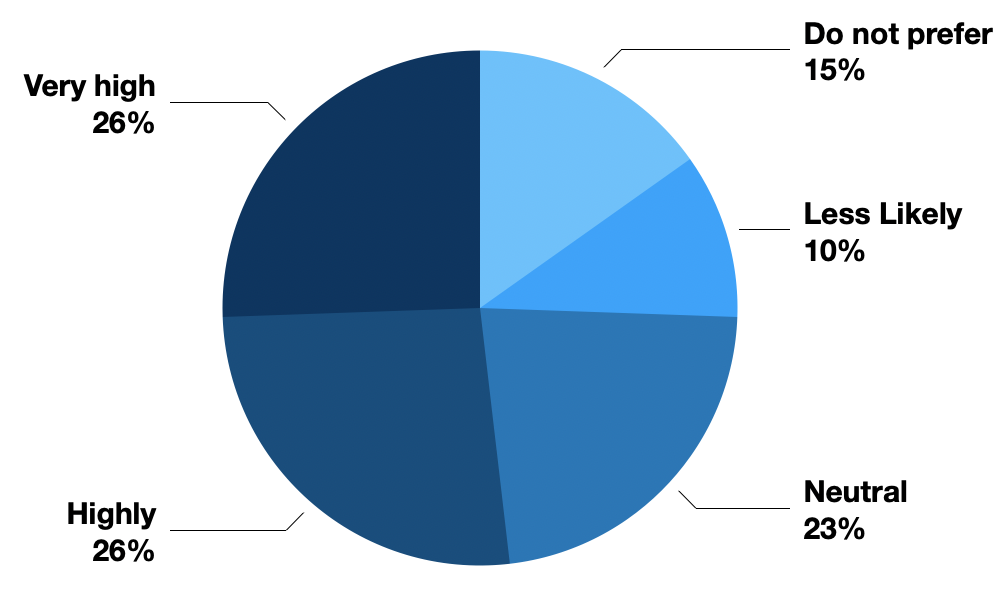}
  \caption{Personalization preference of Millennials and Generation Z respondents.}
  \Description{Pie chart showing the reuslts for the personalization preference of Millennials and Generation Z respondents.}
  \label{fig:preference-content-personalization}
\end{figure}

When asked if they like seeing personalized content recommendations while visiting an online magazine or news media website, 52\% of Millennials/Gen Z responded that their preference for personalized content is very high, or high (see Figure~\ref{fig:preference-content-personalization}), while this percentage is of 30\% in older age groups.

\subsection*{Privacy, a recurring concern}
\label{sub:privacy-a-recurring-concern}
We asked people whether data privacy is a concern to them and, if so, what aspect(s) of it worries them.

\begin{figure}[h]
  \centering
  \includegraphics[width=0.72\linewidth]{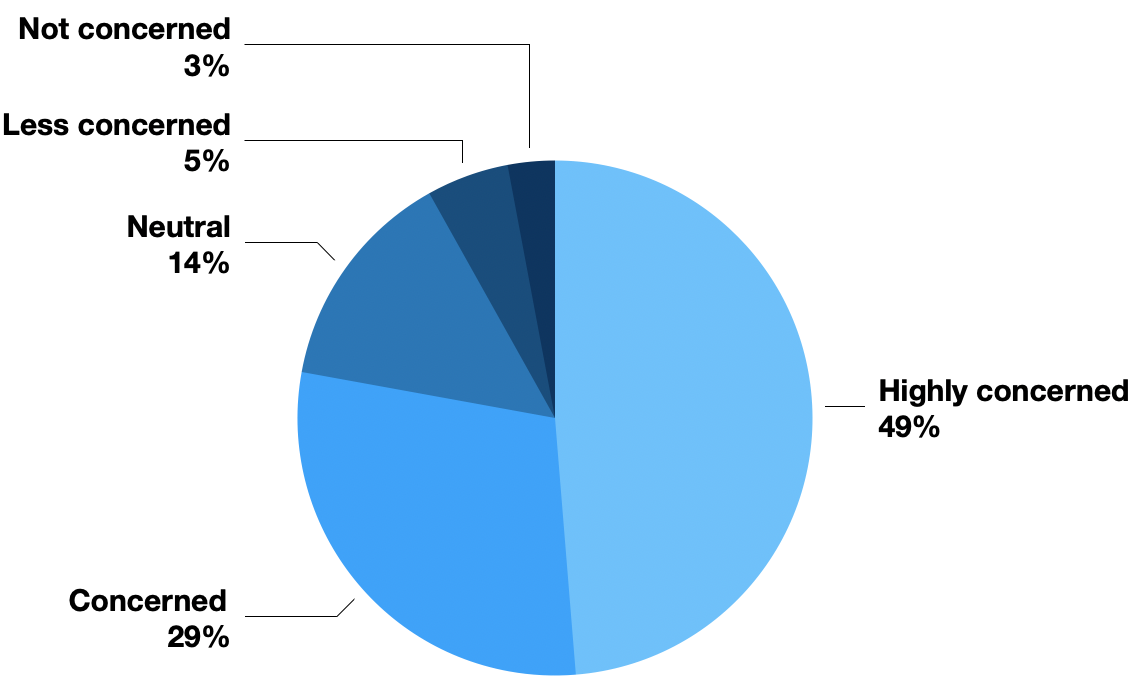}
  \caption{Readers' thoughts on privacy online.}
  \Description{Pie chart showing the results for readers' thoughts on privacy online.}
  \label{fig:concerned-of-privacy}
\end{figure}

The answers reveal that personal data privacy remains a common concern, with 78\% of the respondents worried about their privacy on digital platforms (see Figure~\ref{fig:concerned-of-privacy}). The major concerns were that personal data were being collected by apps and websites and being sold/shared to third party apps.

\subsection*{More data than necessary collected}
\label{sub:more-data-than-necessary-collected}
When asked if they thought companies collect more or less data from them than required to offer a good personalization experience, our respondents felt that more data were collected than necessary in the name of personalization.

Over 63\% of the respondents believed that more than enough data were collected while only 10\% agreed that enough data were collected for personalization.

\begin{figure}[h]
  \centering
  \includegraphics[width=0.5\linewidth]{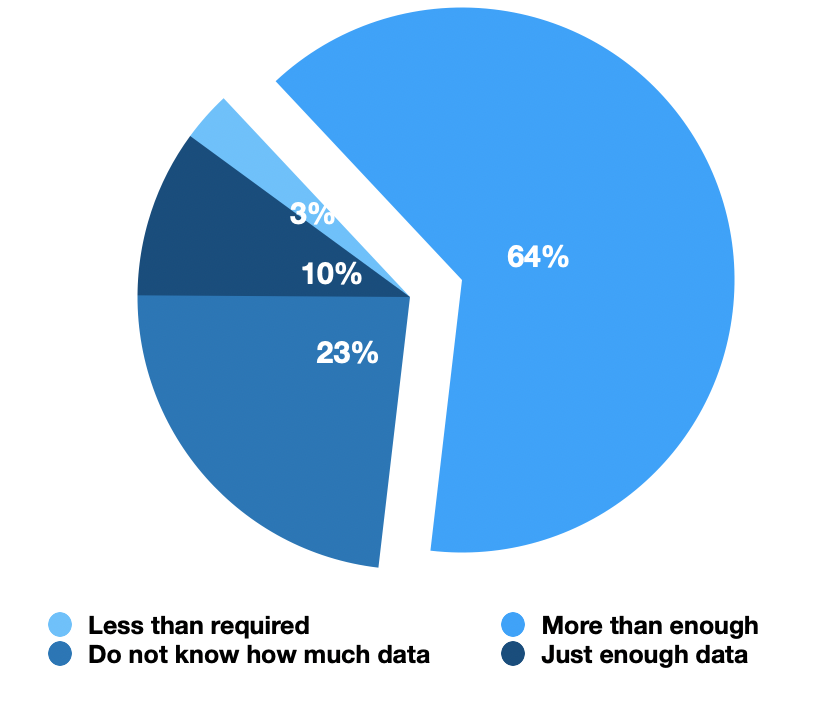}
  \caption{What readers think about data used in the name of personalization.}
  \Description{Pie chart showing the results for what readers think about data used in the name of personalization.}
  \label{fig:how-much-data-required-for-personalization}
\end{figure}

23\% did not know how much data were being collected, which leads us to believe that there is little transparency in the relationship between data collection and its use for personalization (see Figure~\ref{fig:how-much-data-required-for-personalization}).

\subsection*{Stop stalking}
\label{sub:stop-stalking}
Cross-website cookie usage was identified as a common privacy concern among many respondents because of targeted advertisements on social media based on browsing patterns. Another common concern was being asked for mandatory sign-ins or unnecessary personal information. We asked readers about a time they felt their privacy was violated and most of the answers centered around targeted ads.

Examples of anonymous responses received from our participants:

\begin{quote}
\emph{``One search for a product results in me being pestered with ads for the same product in all web sites I visit''}
\end{quote}

\begin{quote}
  \emph{``My search results from an e-commerce site shows up on other websites''}
\end{quote}

\begin{quote}
  \emph{``Mostly the ads whenever I search something on chrome appears on my ads, it’s scary like someone is stalking me always''}
\end{quote}

\begin{quote}
  \emph{``Companies need to act ethically, but for that to happen, regulations must be put in place by governments. A balance absolutely must be found between innovation and personal privacy.''} -- Anonymous respondent.
\end{quote}

\subsection*{Ad-blockers to the rescue}
\label{sub:ad-blockers-to-the-rescue}
\begin{figure}[h]
  \centering
  \includegraphics[width=0.618\linewidth]{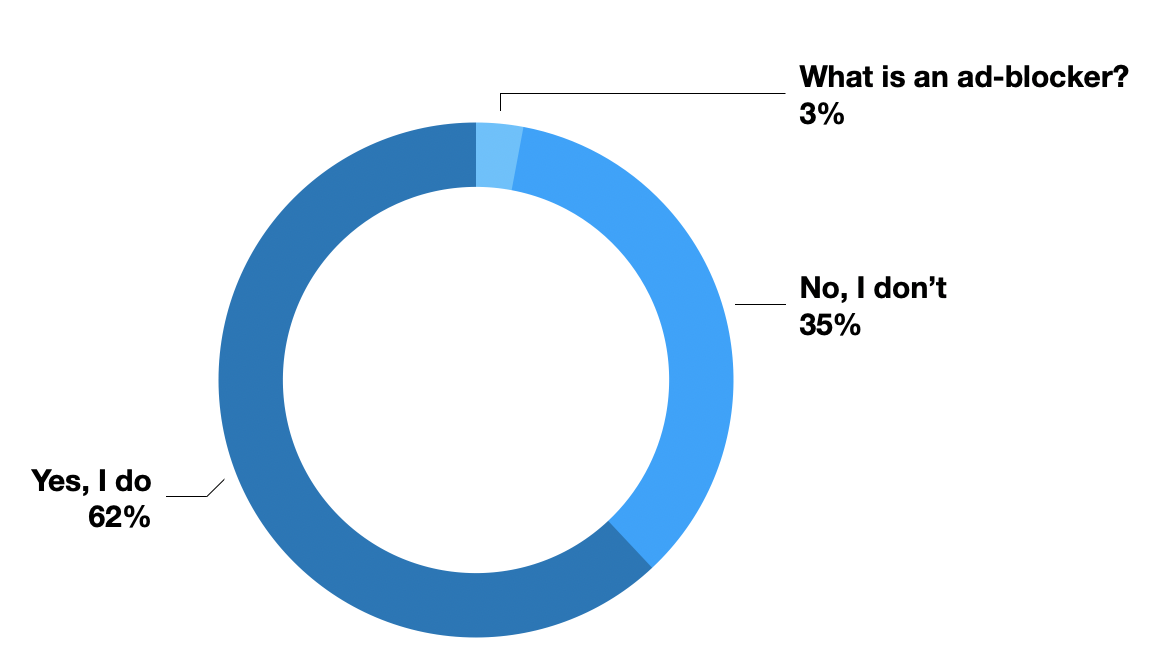}
  \caption{Usage of ad-blockers.}
  \Description{Pie chart showing the reuslts for the usage of ad-blockers.}
  \label{fig:usage-of-ad-blockers}
\end{figure}

When asked if they run Ad blockers in their browsers, 62\% of the respondents agreed to do so, as well as other tools such as Tor or VPN softwares to preserve privacy (see Figure~\ref{fig:usage-of-ad-blockers}).

\subsection*{Key Takeaways}
\label{sec:key-takeaways}
We can summarize the key takeaways as follows:

\begin{itemize}
  \item Millennials and Gen Z readers favor personalization while older groups are still wary of it.
  \item Privacy online is a concern across all age groups. Not only on older age brackets.
  \item There is no transparency in how the data collected are used for personalization and that more data are collected in the name of personalization.
  \item People use their own tools to protect themselves online like ad-blockers, VPN and privacy preserving browsers.
  \item There is a need to reach a balance between personalization experiences and privacy to reap the benefit of both.
\end{itemize}

\section{Conclusion}
\label{sec:conclusion}
People should hold the power in the tech industry, particularly in deciding what data they want to share and how that data are to be used.

With digitalization, publishers and brands have realized that they need to become customer centric with their offerings and move away from ad-based monetization that targets user information and cookies. Building trust and transparency have shown to help publishers increase customer loyalty. Regulations such as GDPR and CCPA have made indiscriminate collection dangerous for companies, forcing them to stop or restrict services in some cases. The need to offer personalized recommendations, however, is still on the rise.

As our research shows, people are becoming more aware of the need to safeguard their personal data from being collected in the name of personalization, and of the dangers and implications of exposure. Readers enjoy personalization but expect an experience where they need not worry about their data being misused.

Techniques like Federated Learning of Cohorts (FLoC) do not require personal information but use cohort behavior analysis to personalize content. On-device machine learning can enable personalization while giving the power to users allowing all processing on the user's device. This opens news doors for digital publishers, for example, to engage with their readers while building trust and adhering to privacy regulations.


\bibliographystyle{ACM-Reference-Format}
\bibliography{biblio}


\begin{thebibliography}{00}


\ifx \showCODEN    \undefined \def \showCODEN     #1{\unskip}     \fi
\ifx \showDOI      \undefined \def \showDOI       #1{#1}\fi
\ifx \showISBNx    \undefined \def \showISBNx     #1{\unskip}     \fi
\ifx \showISBNxiii \undefined \def \showISBNxiii  #1{\unskip}     \fi
\ifx \showISSN     \undefined \def \showISSN      #1{\unskip}     \fi
\ifx \showLCCN     \undefined \def \showLCCN      #1{\unskip}     \fi
\ifx \shownote     \undefined \def \shownote      #1{#1}          \fi
\ifx \showarticletitle \undefined \def \showarticletitle #1{#1}   \fi
\ifx \showURL      \undefined \def \showURL       {\relax}        \fi
\providecommand\bibfield[2]{#2}
\providecommand\bibinfo[2]{#2}
\providecommand\natexlab[1]{#1}
\providecommand\showeprint[2][]{arXiv:#2}

\bibitem[\protect\citeauthoryear{Brasher}{Brasher}{2018}]%
        {atfop}
\bibfield{author}{\bibinfo{person}{E. Brasher}.}
  \bibinfo{year}{2018}\natexlab{}.
\newblock \bibinfo{title}{Addressing the Failure of Anonymization: Guidance
  from the European Union's General Data Protection Regulation}.
\newblock   (\bibinfo{year}{2018}).
\newblock
\showURL{%
\url{https://doi.org/10.7916/cblr.v2018i1.1217}}


\bibitem[\protect\citeauthoryear{{European Parliament and Council of the
  European Union}}{{European Parliament and Council of the European
  Union}}{2016}]%
        {GDPR}
\bibfield{author}{\bibinfo{person}{{European Parliament and Council of the
  European Union}}.} \bibinfo{year}{2016}\natexlab{}.
\newblock \bibinfo{title}{{General Data Protection Regulation – GDPR}}.
\newblock
  \bibinfo{howpublished}{\url{https://eur-lex.europa.eu/eli/reg/2016/679/2016-05-04}}.
    (\bibinfo{year}{2016}).
\newblock
\newblock
\shownote{Accessed: 2021-07-10.}


\bibitem[\protect\citeauthoryear{Hanson, Wei, Veys, Kugler, Strahilevitz, and
  Ur}{Hanson et~al\mbox{.}}{2020}]%
        {10.1145/3313831.3376415}
\bibfield{author}{\bibinfo{person}{Julia Hanson}, \bibinfo{person}{Miranda
  Wei}, \bibinfo{person}{Sophie Veys}, \bibinfo{person}{Matthew Kugler},
  \bibinfo{person}{Lior Strahilevitz}, {and} \bibinfo{person}{Blase Ur}.}
  \bibinfo{year}{2020}\natexlab{}.
\newblock \bibinfo{booktitle}{{\em Taking Data Out of Context to
  Hyper-Personalize Ads: Crowdworkers' Privacy Perceptions and Decisions to
  Disclose Private Information}}.
\newblock \bibinfo{publisher}{Association for Computing Machinery},
  \bibinfo{address}{New York, NY, USA}, \bibinfo{pages}{1--13}.
\newblock
\showISBNx{9781450367080}
\showURL{%
\url{https://doi.org/10.1145/3313831.3376415}}


\bibitem[\protect\citeauthoryear{{State of California, USA}}{{State of
  California, USA}}{2018}]%
        {ccpa}
\bibfield{author}{\bibinfo{person}{{State of California, USA}}.}
  \bibinfo{year}{2018}\natexlab{}.
\newblock \bibinfo{title}{{California Consumer Privacy Act – CCPA}}.
\newblock \bibinfo{howpublished}{\url{https://oag.ca.gov/privacy/ccpa}}.
  (\bibinfo{year}{2018}).
\newblock
\newblock
\shownote{Accessed: 2021-07-10.}


\bibitem[\protect\citeauthoryear{{Stuart A. Thompson and Charlie
  Warzel}}{{Stuart A. Thompson and Charlie Warzel}}{2019}]%
        {nyt_privacy}
\bibfield{author}{\bibinfo{person}{{Stuart A. Thompson and Charlie Warzel}}.}
  \bibinfo{year}{2019}\natexlab{}.
\newblock \bibinfo{title}{{Twelve Million Phones, One Dataset, Zero Privacy}}.
\newblock
  \bibinfo{howpublished}{\url{https://www.nytimes.com/interactive/2019/12/19/opinion/location-tracking-cell-phone.html?action=click&module=Opinion&pgtype=Homepage}}.
    (\bibinfo{year}{2019}).
\newblock
\newblock
\shownote{Accessed: 2021-07-13.}


\bibitem[\protect\citeauthoryear{Wadle, Martin, and Ziegler}{Wadle
  et~al\mbox{.}}{2019}]%
        {10.1145/3314183.3323672}
\bibfield{author}{\bibinfo{person}{Lisa-Marie Wadle}, \bibinfo{person}{Noemi
  Martin}, {and} \bibinfo{person}{Daniel Ziegler}.}
  \bibinfo{year}{2019}\natexlab{}.
\newblock \showarticletitle{Privacy and Personalization: The Trade-off between
  Data Disclosure and Personalization Benefit}. In \bibinfo{booktitle}{{\em
  Adjunct Publication of the 27th Conference on User Modeling, Adaptation and
  Personalization}} {\em (\bibinfo{series}{UMAP'19 Adjunct})}.
  \bibinfo{publisher}{Association for Computing Machinery},
  \bibinfo{address}{New York, NY, USA}, \bibinfo{pages}{319--324}.
\newblock
\showISBNx{9781450367110}
\showDOI{%
\url{https://doi.org/10.1145/3314183.3323672}}


\end{thebibliography}


\end{document}